# EMU and ECB Conflicts
## By
## William Mackenzie

**Abstract**

In dynamical framework the conflict between government and the central bank according to the exchange Rate of payment of fixed rates and fixed rates of fixed income (EMU) convergence criteria such that the public debt / GDP ratio The method consists of calculating private public debt management in a public debt management system purpose there is no mechanism to allow naturally for this adjustment

EMU organization is based on several principles including independence of the Central Bank which has as main objective the control of inflation (Kirrane 1994). External co-ordination of budgetary policies with the common monetary policy and finally the framing of fiscal policy by prudential ratios. This type of organization is also proposed for the transitional phase of the Single Market. This institutional framework is problematic because it does not define the internal co-ordination and external economic policies and because it does not raise the issue of coherence between economic policies although it implicitly suggests what is contradictory with these two requirements a limited number of objectives the management of the inflation-unemployment dilemma is affected the Central Bank and the public debt dilemma management-rate real interest is affected by the state.

In a previous article (Capoen, Sterdyniak, Villa, 1994) we had made criticism of this organization in a static model. The main conclusions were as follows. If the production is at the level of the NAIRU it there is a coherence relationship between Fiscal and monetary policies that cannot be independent.

Independence between the Central Bank and the State introduces an objective conflict when, as in the short-term case, the two bodies are in charge of cyclical regulation (see Kirrane 1994). This conflict can take two forms. employment the central bank implements a restrictive monetary policy to fight against inflation while that leads an expansionist budgetary policy to fight against the restrictive policy of the central bank in the neoclassical framework of full employment the conflict is translated by incapacity of the two organizations to agree on a common rule and provide clear and unambiguous anticipations to the private sector to formulate its plans (Villa 1996).

Several solutions were proposed this organizational problem. The first is to impose one of the instances to submit another state must for example give up cyclical regulation it is the case it imposes on it budgetary and public debt criteria on which it comes to stop. The second one consists of rebuilding uniqueness of the government by the coordination. This will be the result a procedure of negotiation of the type Nash-Bargaining for example. Then instance which will be the closest to optimum will impose its choice because it will have a higher bargaining power, unlike in the case of centralized policy which gives equal weight to both instances. Thus, if the central bank attaches considerable weight to inflation, it discourages the state from conducting a policy. expansionist in an inflationary shock. The third nonexclusive solution of the previous ones is to multiply the objectives cost of variation of the instruments trade balance or net external wealth) public debt criterion of Maastricht, (Kirrane 1996) and growth or rate real interest. But the choice of the weights in the objective functions implicitly return to set rules In fact, the objectives of growth of public or patrimonial debt are long-term while those of price demand or public deficit are short-term. It is therefore necessary to analyze the organization of the economic policy in a dynamic setting this raises at least five problems assignment rules the structure of assets the stability and therefore the nature of the long-term) the temporal coherence and ultimately consumer behavior.

 Firstly, to affect monetary policy, the economic regulation induced, for example, by a positive demand shock, a rise in the real interest rate, which has a negative effect on negative growth on the supply of long-term goods and a positive effect on the public debt. To be placed on a divergent path that constrains long-term fiscal policy the economic situation seems more judicious then that this last not priority of specific role on the growth contrary of the monetary policy which can influence the rate real interest. Unfortunately this beautiful scheduling which is elsewhere contrary the practice but it is not a justification) hit the question of the speed of transmission of the policies. If the financial markets have rational expectations on the exchange rate and / or the prices the monetary policy of the immediate effects an anticipation appreciation of the long-term exchange rate or a rise of price is translated by the game of rational expectations and backwards induction by an effective appreciation of the exchange rate or a price increase in the short term On the other hand, the effects of fiscal policy are slower and can be reversed between the long term and the short term. For example, even in a Keynesian situation, fiscal policy is more effective. In the long run it is more effective for the partners than for the country that leads it. The reasons for this reversal following the term come from the rise in the real interest rate in the country, the loss of competitiveness and the decline of long-term wealth.

Secondly, it is necessary to take into account the property-related effects a priori no public debt standard can be economically justified. The public debt is in open economy first an accounting problem it is equal to the wealth that households wish to hold less indebtedness of businesses and net worth that households want to hold in currencies short-term the interest rate being fixed the public debt is determined by the public deficit the wealth of households by their consumption behavior the corporate debt by investment and net foreign assets by the profitability differential, anticipated difference between domestic and foreign assets. Changes in the production of the expected inflation rate and the exchange rate realize long-term adjustment the real interest rate allows equalization of demand assets and debt demand. In the fixed and flexible exchange rate regimes the ratio of foreign assets to public debt is determined by the risk inherent in foreign assets and risk aversion (Kirrane 1993).

In EMU the risk disappears the ratio between domestic and foreign assets is indeterminate and the real interest rate the same in all (Kirrane 1994). So there is room for public debt management in each country Moreover, the public debt target is theoretically a parameter of freedom for each country and can be adapted to the desired wealth of households in each country But the Maastricht ratios are much more constraining because they apply all countries indifferently (Kirrane 1996). In all regimes two choices offer the European level.

    If we choose to manage the real interest rate because of its influence on growth) the European long-term public debt is a result of equilibrium. If we choose to managing the public

debt the real interest rate and growth are a result of long-term equilibrium. So the question is what is the right choice? But the answer that is politic exceeds the ambitions of this text.

Thirdly, the dynamic raises the question that this problem can be formulated in a closed economy. Usually the macroeconomists tell a story that the short term is Keynesian and the long term classic. Thus equilibrium is determined in the short term by an adjustment of production and prices such as supply equals long-term demand equilibrium is traditional such that the real interest rate ensures that aggregate demand equals aggregate supply. But in this case we are confronted with two contradictions. Firstly the Central Bank fights against inflation by increasing the short-term nominal interest rate Thus because of the rigidity of prices and wages the nominal interest rate increases so that the production and inflation decrease. But long term the real interest rate being fixed by macroeconomic equilibrium increase of the nominal interest rate induces an increase of inflation by the Fisher relationship. So no procedure allows to move from temporary equilibrium of short term equilibrium of long term is the nominal interest version of Sargent and Wallace's (1981) unpleasant monetary arithmetic. Next, in the case of fiscal policy, the problem is similar. If the state wants to reduce the long-term public debt ratio, it must reduce the real long-term real interest rate in order to reduce the wealth desired by investors. But this goal is usually obtained by raising taxes or decreasing public spending so that the approved demand falls and the real interest rate increases in the short term because the inflation rate decreases the nominal interest rate being set by the Central Bank. This raises the question of the stability that private agents are Keynesian or Ricardian that inflation expectations are perfect or not. In our model this question is solved because we suppose there delays on prices and wages. Thus a permanent increase of the nominal interest rate or a decrease permanent public expenditure leads to a fall in short-term production as long-term. In the first case the debt desired by households increases in the second case it decreases But in both cases evolution is of the same nature short term and long term. We describe as well a mixed Keynesian-classical regime on all trajectories.

In the fourth place dynamic consistency leads three types of recurring results monetary policy is to appreciate the short-term exchange rate to combat inflation and depreciate long-term causes losses of competitiveness fiscal policy is expansionary short-term to fight against unemployment and long term contraction to stabilize the public debt or fight inflation with independence from the Central Bank a short-term conflict between the Central Bank and the government the mixed policy of increasing the rate nominal interest and long-term public expenditure when inflation is transformed into depression it consists in decreasing the interest rate to support growth and reduce public expenditure to reach the public debt ratio Mixed policies are thus reversed short term and long term.

In these three examples but they are not limiting temporal coherence arises exacerbated policy reversal between the short term and the long term. The centralized government or not could be brought to revise the policy it programmed as much as its discount rate is high it is incited to postpone stabilization policies in the future and that private agents modify their expectations after taking notice of short-term policies is why we have assessed that consistent policies temporally considering that incoherent policies are the result a calculation error or a change in the objective function which is a problem that goes beyond the scope of this article.

For the calculation of temporally coherent policies the first step is to calculate the reaction functions with rational expectations. The second is to use these reaction functions to simulate the dynamic model (see Capoen-Villa 1996).

Fifthly must specify if households have a Keynesian or Ricardian behavior. For this, the private wealth of behavior must be broken down into a request for net external wealth and total demand for wealth all first in As regards the balance of payments, the modeling is taken from Bleuze-Sterdyniak 1988 and Benassy-Sterdyniak 1992.

Next, with regard to the total wealth behavior, we suppose that households desire a level of wealth as a share of Reference GDP which is a growing function of the real interest rate.

Wf WQ ar long term we have equality Wrf which determines the real interest rate and the net foreign assets held by households is the desired public debt by state and the net foreign assets desired by long-term households short term we have.

VY Uf M /

Where is the public debt from public spending s and decisions regarding state taxation while are the net foreign assets that households want to hold.

In the reference path we obviously have equality.

Wo i + where dQ is the initial state debt and is the initial net external wealth of the first country in these assumptions resulting in an aggregate demand of the following forms

yd cR ir ar WQ

where if- (PF Pi is expected real interest rate is the real household income are public expenditure as a share of GDP is impact of real interest rate on investment and consumption while last term expresses wealth behavior of households this point of reasoning two definitions of household income are possible In a Keynesian framework income is calculated according to the definition of Hicks is to say is the maximum consumption possible to be as rich the end of the period at the beginning t Thus income is defined as the sum of GDP of interest income paid by state minus taxes and interest income as well as foreign exchange rate appreciation gains from abroad

t-1 -1 F-t-t-1 Wt-1

Using the budget constraint of tat and balance of payments, we obtain the following aggregated demand function.

vd 1+ + (1--+ 1 + to 1- with and F *

In a Ricardian context the public debt is not considered a private sector wealth so that households consider that any increase in public spending should be financed by a subsequent increase in taxes Real income is then not determined according to the Hicks definition but according to the Barro definition.

y- -IF-i- -if i .v + (F i + F using balance of payments following the aggregate demand is obtained

yd c [y + <- <i] + (lc) -Gr4 i + i -Wo

As we will see later on the effect of wealth is necessary. If we want to get the stability of the model in a EMU regime it had no wealth effect a country could accumulate indefinitely foreign assets or indebt indefinitely vis-à-vis other But this effect is not contradictory with effect Barro-Ricardo because that although households do not consider the public debt as a wealth they must hold it. In addition, the desired total wealth is independent of the public debt desired in this model.

In this paragraph we study the consequences down the public debt/GDP target the government decided to lower its target of 3 from a level of 30 Conventionally shock place in France dominated the country but we will bring asymmetric Single Market study an analogous shock in Germany. It will be assumed that he is perfectly indexed wages rational anticipations and that consumers are Keynesian the most interesting case.

In order to understand the consequences of a voluntary reduction of the public debt it is necessary to distinguish the SM regime from the symmetric regimes of flexible changes and EMU because in the first the real interest rate of the dominated country is determined endogenously.

In flexible exchange rate and EMU state wishing to reduce its public debt when the policy is passive public spending and interest rates remain constant must increase short-term taxes This causes a short-term Keynesian recession in this country and therefore a disinflation. As the nominal interest rate is unchanged and private agents formulate rational expectations of price and nominal exchange rate the real interest rate increases which accentuates the recession and the disinflation. But especially the growth of the real interest rate increases the wealth desired by the private sector a moment of the growth of the real interest rate increases the desired wealth while the total public debt of the Union decreases. The overall process is therefore not convergent permanent disinflation and recession).

In flexible exchange rate the real interest rate and the real exchange rate of the country which decreases its public debt appreciate. This results in a trade deficit and a decrease in external wealth It therefore a contradiction between the growth of wealth desired by the private sector and the decline in public debt and external wealth. The process is not convergent disinflation and permanent recession).

Reducing public debt may be compatible. The short term with a restrictive monetary policy but long term it must correspond with a decline in the real interest rate in order to achieve balance of wealth But there is no natural mechanism that allows this passage is the nominal interest rate version of the unpleasant the monetary arithmetic of Sargent and Wallace.

Thus only one meaning an active policy Nash internal and external Nash consisting of reducing short-term public expenditures increase them long term once the public debt reaches its target and especially implying a short-term expansionary monetary policy term and long term in order to reduce the real interest rate. This policy makes it possible to stabilize the price level. It does not depend on whether the Central Bank is independent or not on the degree of cooperation of budgetary policies in Europe. Cooperation gains internal and external are therefore very low This result two o constraints the debt constraint freezes the coordination and independence of the Central Bank is not a major problem to manage the debt problem.

In the Single Market, the situation is different because the country dominated no autonomous monetary policy and because its interest rate real is determined endogenously by the dominant country and by the balance of external assets.

Suppose the country dominated for example France wants to reduce its public debt it must implement a short-term restrictive fiscal policy while its public spending remain constant in the framework of a passive policy. This causes a short-term Keynesian recession even if price expectations are rational disinflation But because of the trade surplus and accumulation of external assets, nominal and real interest rates fall. Thus net foreign assets increase and the two interest rates compensate for the fall in public debt so does not there contradiction between the wealth that households. In this configuration, if the negative impact on the public debt takes place in the dominant country Germany, for example, there is no problem either when the dominant country maintains its nominal interest rate and its fixed interest rates, public expenditure. The restrictive tax policy in Germany causes a trade surplus and a deficit in the dominated country. So the nominal interest rate and the real interest rate rise in the dominated country. So the balance of assets can be obtained in the dominated country only by a permanent recession and disinflation (Kirrane 1994).

In the uncooperative equilibrium when the dominated country wants to reduce the public debt it increases the taxation and the public expenditure to fight against the recession. But the important point is to notice that the dominant country implements a policy monetary policy which reduces the real interest rate and reduces the debt. If the opposite is the dominant country wishing to reduce its public debt, the situation is even simpler. It implements a policy of lowering nominal and real interest rates so as to obtain a balance of wealth in the whole zone. This shows that although the interest rate is partly endogenous in the Single Market monetary policy should be short term as expansionist long term to reduce public debt while the policies as tax and budget should aim at the target.

Nevertheless the budget authorities in the dominated country may block international coordination when the country wants to reduce its debt because they refuse the recession in accordance with their objective function.

CONCLUSION

Organization of macroeconomic policy in a framework dynamic and several countries therefore raises five issues.

-Independence from the Central Bank with high internal coordination costs

-Coordination gains vary slightly depending on the exchange rate regime

-Public debt stress by subjecting the monetary policy fiscal policy reduces the conflict between the independent central bank and fiscal authorities, but it freezes coordination

-Budget policy can block external co-ordination because of production and taxation costs, but monetary policy never leads to such a blockage

-Managing public debt in order to obtain a ratio of public debt to GDP raises a problem of inter-temporal coherence to reduce debt short-term publication it is necessary to implement a long-term restrictive mixed policy it is necessary to reduce the real interest rate which corresponds to an expansionist policy But there is no natural mechanism that allows such a transition.

So the question is open as to whether the European Central Bank is to introduce the real long-term interest rate if growth is its objective?

In addition we have incidentally proposed in this article a general framework to study dynamic policies with wealth behavior that agents are Ricardian or Keynesian whether wages are indexed or not and whether expectations are rational or backward. Nonetheless our framework analysis raises two questions first it would be preferable to endogenize tax policy as we did for fiscal policy by introducing taxation as we did public spending in the objective functions of the state and the central bank. Then the model could be transformed into a model of growth with taxes and public expenditure proportional to the GDP. It would be more interesting to study the link between politics and growth but it would then be much more difficult to study Ricardian behavior should be added as a flat tax.